\documentclass[epj]{svjour}

\def\beq{\begin{equation}}
\def\eeq{\end{equation}}
\def\beqar{\begin{eqnarray}}
\def\eeqar{\end{eqnarray}}
\def\barr#1{\begin{array}{#1}}
\def\earr{\end{array}}
\def\bfi{\begin{figure}}
\def\efi{\end{figure}}
\def\btab{\begin{table}}
\def\etab{\end{table}}
\def\bce{\begin{center}}
\def\ece{\end{center}}

\def\text{\textstyle}

\def\De{\Delta}

\def\refeq#1{\mbox{(\ref{#1})}}

\def\citere#1{\mbox{Ref.~\cite{#1}}}
\def\citeres#1{\mbox{Refs.~\cite{#1}}}

\def\mathswitchr#1{\relax\ifmmode{\mathrm{#1}}\else$\mathrm{#1}$\fi}

\newcommand{\PW}{\mathswitchr W}

\def\mathswitch#1{\relax\ifmmode#1\else$#1$\fi}

\def\solid{\raise.9mm\hbox{\protect\rule{1.1cm}{.2mm}}}
\def\dash{\raise.9mm\hbox{\protect\rule{2mm}{.2mm}}\hspace*{1mm}}

\begin{document}
\title{New Predictions for Electroweak $\mathbf{\mathcal{O}(\alpha)}$ Corrections to 
Neutrino--Nucleon Scattering}

\subtitle{Talk given at the EPS conference, Aachen, July 2003}
\author{K.-P.~O.~Diener
\thanks{in collaboration with S.~Dittmaier and W.~Hollik}%
}                     
%
%
\institute{Paul Scherrer Institut, CH-5232 Villigen PSI, Switzerland}
\date{Received: date / Revised version: date}
%
\abstract{We calculate the ${\cal O}(\alpha)$ electroweak corrections to charged-
and neutral-current deep-inelastic neutrino scattering off an isoscalar target.
The full one-loop-corrected cross sections, including hard photonic corrections, 
are evaluated and compared to an earlier result which is the basis of
the NuTeV analysis.
In particular, we compare results that differ in input-parameter 
scheme, treatment of real photon radiation and factorization scheme.
The associated shifts in the theoretical prediction for the ratio of neutral- and
charged-current cross sections can be larger than the experimental accuracy of 
the NuTeV result. 
This work is described in more detail in a recently published paper\cite{Diener:2003ss}.
\PACS{
      {12.15.-y}{Electroweak interactions} \and  {12.15.Lk}{Electroweak radiative corrections} 
        \and
      {25.30.Pt}{Neutrino scattering} 
     } 
} 
\maketitle
\section{Introduction}
\label{intro}
Deep-inelastic neutrino scattering has been analyzed in the
NuTeV experiment \cite{Zeller:2001hh} with a rather high precision.
In detail, the neutral- (NC) to charged-current (CC) cross-section 
ratios~\cite{LlewellynSmith:ie}
\[
R^\nu = \frac{\sigma_{NC}^\nu(\nu_\mu N\to\nu_\mu X)}
{\sigma_{CC}^\nu(\nu_\mu N\to\mu^- X)}, 
\;
R^{\bar\nu} = \frac{\sigma_{NC}^{\bar\nu}(\bar\nu_\mu N\to\bar\nu_\mu X)}%
{\sigma_{CC}^{\bar\nu}(\bar\nu_\mu N\to\mu^+ X)} 
\]
have been measured to an accuracy of about 0.2\% and 0.4\%, respectively.
In addition, the quantity
\[
R^- = \frac{\sigma_{NC}^\nu(\nu_\mu N\to\nu_\mu X)
-\sigma_{NC}^{\bar\nu}(\bar\nu_\mu N\to\bar\nu_\mu X)}
{\sigma_{CC}^\nu(\nu_\mu N\to\mu^- X)-\sigma_{CC}^{\bar\nu}(\bar\nu_\mu N\to\mu^+ X)},
\]
as proposed by Paschos and Wolfenstein \cite{Paschos:1972kj},
has been considered.
As a central result, the NuTeV collaboration has translated their
measurements of $R^{\nu/\bar \nu}$ and $R^-$ 
into values for the on-shell weak mixing angle,
$\sin^2\theta_W=1-m_W^2/m_Z^2$, 
which can be viewed as independent (but
rather indirect) determinations of the W- to Z-boson mass ratio.
The NuTeV result on $\sin^2\theta_W$ is, however, about $3 \sigma$ away from the
result obtained from the global fit \cite{Grunewald:2003ij}
of the Standard Model (SM) to the electroweak precision data.

It was pointed out~\cite{McFarland:2003jw}
that the inclusion of electroweak radiative corrections,
which influences the result significantly, is based on a single
calculation \cite{Bardin:1986bc} only
\footnote{Electroweak radiative corrections were also investigated in
Refs.~\cite{Marciano:pb}, where in the numerical evaluation several
approximations were made. Although the input used in Refs.~\cite{Marciano:pb} 
is obsolete, the prediction made in these references is relatively close 
to the results of the analysis presented here.}
and that a careful recalculation of these corrections would be desirable. 
In this work we summarize a recent publication~\cite{Diener:2003ss}
describing such a calculation of the ${\mathcal O}(\alpha)$ 
electroweak corrections to NC and CC deep-inelastic 
neutrino scattering off an isoscalar target. 
Apart from a different set of parton densities and input parameters 
the most important difference between our calculation and the result of 
Ref.~\cite{Bardin:1986bc} lies in the treatment of mass
singularities due to collinear radiation of a photon from external 
charged particles. 


\section{Lowest-Order Results}
\label{sec:LO}

We consider the NC and CC parton processes
\begin{eqnarray}
\mbox{NC:} \quad \nu_\mu(p_l) + q(p_q) &\;\to\;& \nu_\mu(k_l) + q(k_q), 
\label{eq:ncprocesses}
\\
\mbox{CC:} \quad \nu_\mu(p_l) + q(p_q) &\;\to\;& \mu^-(k_l) + q'(k_q),
\label{eq:ccprocesses}
\end{eqnarray}
where in Eq.~(\ref{eq:ncprocesses}) the generic label $q$ stands for all
light quark and antiquark flavours (including charm) and in Eq.~(\ref{eq:ccprocesses}) 
it stands for the quarks $\mathrm{d},\mathrm{s},\bar\mathrm{u},\bar\mathrm{c}$ 
($q'$ represents all CKM-allowed light final-state quarks).
Additionally we consider the processes with all particles replaced by their antiparticles.
With the usual Bjorken scaling variable $x$,
neglecting all fermion masses where it is consistently possible, the 
squared partonic centre of mass energy $s$ is given by
\[
s = 4E^2 = 2x M_N E_\nu^\mathrm{LAB},
\]
up to terms of higher order in the nucleon mass.
Hadronic cross sections are obtained by convoluting
parton-level cross sections with iso-averaged 
parton density functions (PDFs) which account for the isoscalar composition
of the nuclear target.

At leading order the approximate relation
\begin{equation}
R^\nu \sim \frac{1}{2}-\sin^2\theta_\PW+\frac{20}{27}\sin^4\theta_\PW
\label{eq:sw2}
\end{equation}
and similar relations for $R^{\bar \nu}$ and $R^-$ 
permit the translation of an experimental determination 
of the quantities $R^{\nu/\bar \nu}$, $R^-$ 
into an indirect measurement of $\sin^2 \theta_W$.

\section{Higher-Order Corrections}
\label{sec:NLO}

The inclusion of quantum effects in the theoretical
prediction can be incorporated in Eq.~\refeq{eq:sw2} in terms of a 
small variation, ultimately relating the relative higher-order 
corrections to the NC and CC cross sections
to a shift in the predicted value of the weak mixing angle:
\begin{equation}
\Delta \sin^2 \theta_W =
\frac{\frac{1}{2}-\sin^2\theta_W+\frac{20}{27}\sin^4\theta_W}
{1-\frac{40}{27}\sin^2\theta_W}
\left( \frac{\delta\sigma^\nu_{NC}}{\sigma^\nu_{NC}} 
	- \frac{\delta\sigma^\nu_{CC}}{\sigma^\nu_{CC}} \right).
\label{eq:dsw2}
\end{equation}

All parts of our calculation of $\mathcal{O}(\alpha)$ corrections 
to the leading order matrix elements have been performed in two
independent ways, resulting in two completely independent computer
codes. Both loop calculations are carried out in `t~Hooft--Feynman gauge
and are based on the standard techniques for
one-loop integrations as, e.g., described in 
\citeres{Denner:1993kt,'tHooft:1979xw}. Ultraviolet divergences
are treated in dimensional regularization and eliminated using the
on-shell renormalization scheme \cite{Denner:1993kt,Bohm:rj} in the
formulation of \citere{Denner:1993kt}.
Infrared (i.e.\ soft and collinear) singularities are regularized
by an infinitesimal photon mass and small fermion masses.
The artificial photon-mass dependence of the virtual and (soft) real 
corrections cancels in the sum of both contributions, according to
Bloch and Nordsieck \cite{Bloch:1937pw}. 

The calculation of virtual and real corrections is by now a 
standard exercise and was in part performed with the help of 
suitable computer algebra programs like {\sc FeynArts} \cite{Kublbeck:1990xc},
{\sc FormCalc} \cite{Hahn:1998yk} and {\sc FeynCalc}
\cite{Mertig:1991an} and in part carried out with independent computer-algebra routines or
with recourse to related published work~\cite{Dittmaier:2001ay}.
Among the virtual corrections, there are two contributions to the
NC processes that become numerically delicate in the limit of small
momentum transfer in the Mandelstam variable $t$:
the $\gamma\nu_\mu\bar\nu_\mu$ vertex correction and the $\gamma Z$
mixing self-energy. 
The limit \mbox{$t \to 0$} is physically well-defined in both cases,
but the numerical treatment of the corresponding amplitudes deserves 
some care to ensure proper cancellation of powers of $t$ in the 
form factors against the $t$-channel photon propagator.

Initial-state mass singularities due to collinear photon radiation
were subtracted from the real corrections with a suitably defined
$\overline{\mbox{MS}}$ counterterm 
(see, e.g., Refs.~\cite{Dittmaier:2001ay,Baur:1999kt}), 
as it is standard procedure in perturbative QCD.
For the numerical evaluation of the cross sections, however, we adopted 
leading order CTEQ4L~\cite{Lai:1997mg} parton densities, 
which is formally inconsistent with the aforementioned subtraction.
Nevertheless, this procedure is acceptable as a full incorporation of 
$\mathcal{O}(\alpha)$ effects in the DGLAP evolution of PDFs  
and a corresponding fit to experimental data has not yet been performed
and the overall effect of QED initial-state collinear radiation largely
cancels in the radiative corrections to the quantities $R^{\nu,\bar \nu}$ 
and $R^-$ (see Eq.~\refeq{eq:dsw2}).
It should be mentioned that our method of initial-state mass 
factorization is fundamentally different
from the technique employed in \citere{Bardin:1986bc} (called BD below), 
where the initial-state mass dependence is left unsubtracted 
and \mbox{$m_q = x \, m_N$}, the scaled nucleon mass,
is chosen for the initial-state mass value.

There are also $\alpha\ln m_{q'}$ and (in the CC case)
$\alpha\ln m_\mu$ terms from final-state radiation.
According to the Ki\-no\-shi\-ta--Lee--Nauenberg (KLN) theorem \cite{KLN},
these terms drop out if the final state is treated sufficiently inclusive,
i.e.\ if the cones for quasi-collinear photon emission around the charged
outgoing fermions are fully integrated over.
This condition is, in general, not fulfilled if phase-space cuts at the
parton level are applied. 
In the NuTeV analysis an event is discarded unless the
energy deposited in the calorimeter lies within certain bounds,
which imposes a cut on the final-state particles' energies.
We have implemented this final-state energy cut in three different 
ways in our theoretical analysis, namely by imposing it either (1) 
on the final-state quark alone, (2) on the final-state quark and 
final-state real photon energy added together or (3) 
on the final-state quark alone except if the real photon 
is emitted within a cone of $5^\circ$ (in the 
laboratory frame) around the final-state quark.
For the first procedure the KLN theorem predicts a
dependence of our numerical results on the final-state quark mass,
in the second case there is some residual dependence 
on the muon mass.
The third method, imposing the cut on the recombined quark and photon
energy, renders our results independent of any final-state mass.
Although this seems to be a strong argument in favor of the 
recombination technique, the cut on the
sum of hadronic and photonic energies seems
to be most appropriate for a fixed-target experiment where all but the
energy of neutrinos and muons is deposited in a calorimeter.
To demonstrate the effect of different implementations  of 
the final-state energy cut we present numerical results for all three
procedures described above.

\section{Numerical Results}
\label{sec:numerics}

The authors of \citere{Bardin:1986bc} advocate the on-shell scheme with
$G_F, m_Z, m_W, m_H$ and the fermion masses as independent input parameters.
To compare our results,
we adopt their numerical values for $G_F$, $\alpha(0)$, 
the Z- and Higgs-boson mass, the top-quark mass
and the electroweak mixing angle $\sin^2 \theta_W$. 
We supplement these parameters by the missing
quark and lepton masses from Refs.~\cite{Sarantakos:1982bp,Roos:sd} 
which are quoted within \citere{Bardin:1986bc}.
\begin{table}
\begin{tabular}{ccccc}
	  \multicolumn{2}{l}{result of \citere{Bardin:1986bc}} & \multicolumn{3}{c}{$-114$} \\ \hline \hline
	  input &  factorization  & \multicolumn{3}{c}{final-state energy cut:} \\
	  parameters: & scheme:      & \hspace{2mm} (1) \hspace{2mm} & \hspace{2mm} (2) 
		\hspace{2mm}& \hspace{2mm}(3) \\ \hline
	  $G_F, \sin^2 \theta_W$ & $\overline{\mbox{MS}}$ & $-90$  & $-130$ & $-94$ \\
	  $G_F, \alpha(0)$       & $\overline{\mbox{MS}}$ & $-95$  & $-132$ & $-99$ \\
	  $G_F, \sin^2 \theta_W$ & BD                     & $-98$  & $-138$ & $-102$ \\
	  $G_F, \alpha(0)$       & BD                     & $-103$ & $-139$ & $-106$ \\
\end{tabular}
\caption{ \label{tab:numres} 
Compilation of results for $\Delta \sin^2 \theta_W \cdot 10^4$ calculated
from Eq.~\refeq{eq:dsw2}.
We compare the prediction of \citere{Bardin:1986bc} with ours in 
different input-parameter and initial-state mass factorization schemes 
(see Section~\ref{sec:NLO} for details) and for
different variants of final-state energy cut (we required $E_\mathrm{fin} > 10\mathrm{GeV}$).
A complete specification of numerical input parameters and more details on
the calculations are found in Ref.~\cite{Diener:2003ss}}
\end{table}
Unfortunately it is not completely clear, whether, 
given this set of input parameters, the W-boson mass used in
\citere{Bardin:1986bc} is calculated from $m_Z$ and $\sin^2 \theta_W$
or by iterative solution of the relation 
\[ m_W^2 \left(1-\frac{m_W^2}{m_Z^2}\right) = 
\frac{\pi\alpha}{\sqrt{2} G_F} \frac{1}{1-\Delta r(\alpha,m_W,m_Z,m_H,m_f)}
\]
from $G_F, \alpha(0)$ and all particle masses except $m_W$. In 
Table~\ref{tab:numres} the two alternative parameterizations are denoted 
``$G_F, \sin^2 \theta_W$'' and ``$G_F, \alpha(0)$'', respectively.

The relevant numerical result of \citere{Bardin:1986bc}
and a compilation of our results for $\Delta \sin^2 \theta_W$ as 
given in Eq.~\refeq{eq:dsw2} for different 
input-parameter- and initial-state mass factorization schemes
are shown in Table~\ref{tab:numres}.
We present numbers for different treatments (labelled 1--3) 
of the final-state energy cut as explained in the previous section.
Of course, this comparison of results can neither prove nor
disprove the correctness of the results of \citere{Bardin:1986bc}.
However, the table 
suggests significant differences in the correction $\De\sin^2\theta_\PW$,
no matter what final-state energy cut or what input-parameter- or
initial-state mass factorization scheme we chose.
In any case, the variations in the corrections that are due to the
different factorization schemes ($\overline{\mbox{MS}}$ versus BD)
and due to different ways of including the final-state photon
in the hadronic energy in the final state can be as large as
the accuracy in the NuTeV experiment, which is about $16 \cdot 10^{-4}$ in
$\sin^2\theta_W$ (if statistical and systematic errors are
combined quadratically).

\section{Conclusions}
\label{sec:conclusions}

A new calculation of electroweak ${\cal O}(\alpha)$
corrections to NC and CC neutrino deep-inelastic
scattering has been presented and compared to an older work~\cite{Bardin:1986bc}. 
The issue of collinear fermion-mass singularities,
have been discussed in detail.
Drawing a comparison to the results of \citere{Bardin:1986bc},
which were used in the NuTeV data analysis, is not straightforward
as the exact parameterization of the input data is not clear.
However, a comparison based on an 
assumption for missing input parameters
seems to suggest significant differences in the
electroweak radiative corrections.
Therefore, an update of the NuTeV analysis seems to be desirable.
We provide a Fortran code for the electroweak radiative corrections
that could be used in this task.

Specifically, our investigation of the factorization-scheme dependence
for initial-state radiation and of different ways to treat photons in the
final state reveals that these effects can be as large as the
$3\sigma$ difference between the NuTeV measurement and the Standard Model
prediction in the on-shell weak mixing angle.

The NuTeV collaboration estimated the theoretical uncertainty due to
missing higher-order effects to 0.00005 and 0.00011 in
$\delta R^\nu$ and $\Delta \sin^2 \theta_W$, respectively.
The results on electroweak corrections presented in this paper
indicate that these numbers might be too optimistic.

\end{document}